\pdfoutput=1
\documentclass[a4paper,12pt]{iopart}
\usepackage{graphicx}
\usepackage{cite}
\usepackage{color}
\usepackage{amsfonts}
\usepackage{srcltx}
\begin{document}
\title{On entropy production in nonequilibrium systems}
\author{ Robert Ziener$^{1}$, Amos Maritan$^{2}$ and Haye Hinrichsen$^{1}$}
\address{$^{1}$Universit\"at W\"urzburg, Fakult\"at f\"ur Physik und Astronomie, Am Hubland, \\ 97074 W\"urzburg, Germany.\\ 
$^{2}$Dipartimento di Fisica 'G. Galilei', CNISM, INFN, Universit\`a di Padova, Via Marzolo 8, 35131 Padova, Italy.
}

\ead{robert.ziener@uni-wuerzburg.de\\ \hspace{11mm}
    maritan@pd.infnf.it \\ \hspace{11mm}
     hinrichsen@physik.uni-wuerzburg.de}
     
\def\d{{\rm d}}

\begin{abstract}
In this paper we discuss the meaning of the Schnakenberg formula for entropy production in non-equilibrium systems. To this end we consider a non-equilibrium system as part of a larger isolated system which includes the environment. We prove that the Schnakenberg formula provides only a lower bound to the actual entropy production in the environment. This is also demonstrated in the simplest example of a three-state clock model.
\end{abstract}

\section{Introduction}

With the development of stochastic thermodynamics~\cite{Sekimoto10,Seifert12,Zhang12,Broeck15} on the basis of nonequilibrium statistical physics~\cite{Prigogine61,Glansdorff71,Groot84} the study of the thermodynamic implications of coarse-graining attracted increasing attention~\cite{Rahav07,Gomez08,Puglisi10,Bulnes11,Bulnes13,Diana14,Mehl12,Leonard13,Altaner12,Horowitz13,Bo14}. In this field Markov jump processes are used to model a large variety of physical systems. Such models possess a discrete set of possible configurations (microstates) $s \in \Omega_{sys}$ and evolve dynamically by spontaneous uncorrelated jumps between the configurations according to certain transition rates $w_{s \to s'}$. Whenever such a jump occurs the system is said to produce an entropy of the amount
\begin{equation}
\label{Schnakenberg}
\Delta S_{env} \;=\; \ln \frac{w_{s \to s'}}{w_{s'\to s}}
\end{equation}
in the environment~\cite{Seifert12}. This formula for entropy production, which goes back to a work by  Schnakenberg~\cite{Schnakenberg76} in 1976, is nowadays used throughout the whole literature (see e.g. Refs.~\cite{tome12,andrieux06,tome15,zia06,zia07}). But where does this formula come from? Does it describe the actually generated entropy outside of the system or does it just have the meaning of a lower bound? The aim of this paper is to shed some light on these fundamental questions.

The original approach taken by Schnakenberg is quite remarkable. He first points out that the master equation
\begin{equation}
\label{masterequation}
\frac{\d}{\d t} P_s(t) = \sum_{s'} P_{s'}(t) w_{s'\to s} - \sum_{s'} P_s(t) w_{s \to s'}
\end{equation}
for the configurational probabilities $P_s(t)$ is formally equivalent to a chemical rate equation
\begin{equation}
\frac{\d}{\d t} [X_i] = \sum_{j} S_{ij} f_j\Bigl([X_1],[X_2],\ldots\Bigr)
\end{equation}
for particle concentrations $[X_i]$ with an appropriately chosen stoichiometric matrix $S_{ij}$ and a linear rate function $f$. Thus, by identifying the probabilities $P_s$ of individual microscopic configurations $s$ with the concentrations $[X_i]$ of different chemical species $X_i$ he created a fictitious chemical reaction which evolves formally in the same way as the original master equation. Building such a chemical system is of course only a thought experiment because in practice a complex system like a gas could easily have more than $10^{10^{200}}$ different microscopic configurations, each of them corresponding to a different chemical species. Nevertheless, such a chemical reaction, if realized, would generate a certain entropy which can be quantified by using standard methods of equilibrium thermodynamics. Schnakenberg suggested that this chemical entropy production, when properly normalized, coincides with the actual entropy production of the non-equilibrium system in its environment. 

Some time ago we suggested that the Schnakenberg entropy production is not an equality but only a lower bound for the actual entropy production in the environment~\cite{kolkata}. This claim was first proven by Esposito~\cite{Esposito85} in a thermodynamic setting and then developed further in the context of hidden entropy production~\cite{Kawaguchi13,chun15}. In this approach the subsystems are always in contact with heat baths which allows one to quantify the corresponding energy, work, heat transfer, and entropy flow.

Here we present an alternative proof which is to a large extent independent of thermodynamic notions. As in previous works we embed the laboratory system into a larger isolated (closed) system which can also be modeled as a Markov process. This environment may be out of equilibrium as well, and neither the system nor the environment are coupled to other external heat baths. Using this setup we argue that that any nonequilibrium system may be thought of as being part of a larger equilibrium system on its way into the stationary state. Although we use the same concept of coarse-graining as in Ref.~\cite{Esposito85}, our approach is solely based on microscopic configurations and transitions without assuming energy conservation. Thus the proof does not invoke the notions of energy, heat, work and temperature, showing that the suggested Schnakenberg inequality is to a large extent independent of the first law of thermodynamics. We also prove that the inequality becomes sharp in the limit of instantaneous equilibration of the environment after each jump in the laboratory system.

The derivation is based on the assumption that the 'total system', consisting of laboratory system and environment, is isolated and can still be described by a Markov jump process. As such, it is expected to relax into a Gibbs state of maximal entropy, and this requires the transition rates between microscopic configurations to be symmetric. This assumption could of course be questioned. As an isolated system, quantum mechanics tells us that the 'total system' evolves unitarily, preserving entropy without spontaneous jumps. This contradiction, which touches the very foundations of statistical mechanics~\cite{Gogolin15} will not be addressed in this paper, we rather \textit{assume} that a description in terms of Markov processes still holds and study the resulting consequences.

The paper is organized as follows. In the next section we introduce notations and outline the strategy of embedding a nonequilibrium system in a larger isolated system. Then in Sect. \ref{sec_num} we compare the Schnakenberg formula and the actual entropy production numerically and demonstrate the embedding in the example of a simple clock model. Finally, in Sect. \ref{sec_analytical} we prove that Schnakenbergs entropy production provides a lower bound to the actual entropy production which becomes sharp in the limit of instant equilibration in the environment.

\section{Embedding a nonequilibrium system in the environment} 

Let us consider a physical system, from now on called \textit{laboratory system}, which is modeled as a classical stochastic Markov process. It is defined by a certain space of classical configurations $s\in \Omega^{sys}$, where we use the symbol '$s$' to remind the reader that this configuration refers to the laboratory system. As before, let us denote by $P_s(t)$ the probability to find the system at time $t$ in the configuration~$s$, normalized by $\sum_{s\in \Omega^{sys}} P_s(t)=1$. In a Markov process this probability distribution evolves according to the master equation~(\ref{masterequation}), which can also be written as
\begin{equation}
\label{masterJ}
\frac{\d}{\d t} P_s(t)  = \sum_{s' \in \Omega^{sys} } \Bigl(J_{s' \to s}(t) - J_{s \to s'}(t)\Bigr)\,,
\end{equation}
where $J_{s \to s'}(t) = P_s(t)w_{s \to s'}$ is the probability current flowing from $s$ to $s'$. For simplicity we shall assume that the system is ergodic. 

According to Kolmogorovs criterion~\cite{Kelly}, which implies detailed balance in the stationary state, the rates of an equilibrium system are known to obey the condition
\begin{equation}
\label{pathindependence}
\prod_\gamma \frac{w_{s_i \to s_{i+1}}}{w_{s_{i+1} \to s_i}} = 1
\end{equation}
for all \textit{closed} paths $\gamma: s_1 \to s_2 \to \ldots \to s_N \to s_1$ in the configuration space. The path independence allows each configuration $s$ to be associated with a dimensionless potential
\begin{equation}
V_s \;=\; V_{s_0} - \sum_{\gamma_{s_0\to s}} \ln\frac{w_{s_i \to s_{i+1}}}{w_{s_{i+1}\to w_s}}\,,
\end{equation}
where $s_0$ is a reference configuration and $V_{s_0}$ its reference potential. As can be verified easily, the stationary probability distribution, normalized by the partition sum, is then given by 
\begin{equation}
P_s^{stat} \;=\; \lim_{t \to \infty}P_s(t) \;=\;\frac{1}{Z} e^{-V_s}\,.
\end{equation}
This stationary state is known to obey detailed balance, meaning that the probability currents cancel pairwise:
\begin{equation}
\Rightarrow \qquad J^{stat}_{s\to s'} = J^{stat}_{s' \to s}
\end{equation}
The form of the potential depends on the way in which the system interacts with the environment. For example, in the canonical ensemble we just have $V_s=\beta E_s$.

A special situation emerges if the system is \textit{isolated}, meaning that it does not interact with other systems or the environment. As discussed in the introduction, we start with the premise that it still makes sense to model such a closed system by a Markov process. Starting from this premise we are forced to assume that all rates of an isolated system have to be symmetric ($w_{s \to s'}=w_{s'\to s}$) since otherwise the Gibbs postulate would be violated. Although this reasoning is made on a classical basis, it resembles the well-known Stinespring theorem~\cite{Stinespring55} in the quantum case, stating that any non-unitary open system can be thought of as being part of a larger unitarily evolving system.

\subsection*{Open system embedded in the environment}
The rates of a genuine non-equilibrium system do not obey the Kolmogorov criterion~(\ref{pathindependence}), meaning that it relaxes into a non-equilibrium steady state (NESS) that violates detailed balance. In the following we argue that we can always think of such a system as being embedded into a larger isolated system, called \textit{total} system. This provides a clear setting for the study of entropy production.

\begin{figure}
\centering\includegraphics[width=90mm]{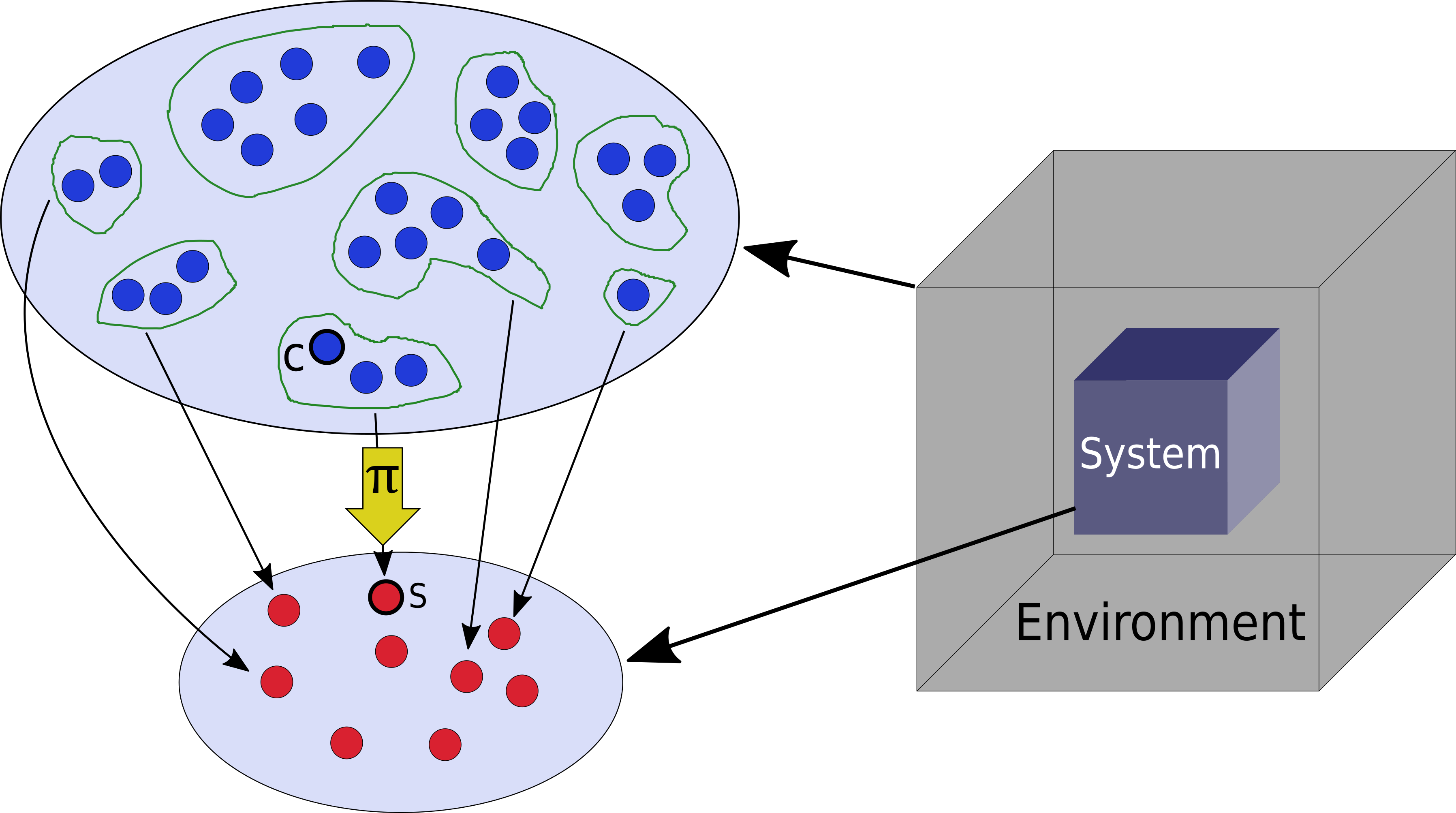}
\caption{Schematic representation of the configuration space of a system embedded in the environment.}
\label{figenvsys}
\end{figure}

Let us assume that this total system (consisting of laboratory system and environment) can be described in the same way as the laboratory system itself, namely, as a Markov process in terms of certain configurations $c \in \Omega^{tot}$, probabilities $P_c(t)$, and time-independent transition rates $w_{c \to c'}$. The configuration space of the total system may be incredibly large and generally inaccessible, but if the total system is assumed to be isolated, hence the corresponding transition rates have to be symmetric:
\begin{equation}
w_{c\to c'} = w_{c'\to c}\,.
\end{equation}
The laboratory system is part of the total system. As such, any configuration $c \in \Omega^{tot}$ of the total system will correspond to a particular configuration $s \in \Omega^{sys}$ in the laboratory system. This mapping
\begin{equation}
\pi: \ c \;\mapsto\; s=\pi(c) 
\end{equation}
is of course not injective, i.e., for a given system configuration $s$ there are usually many possible configurations $c$ of the total system, as sketched in Fig.~\ref{figenvsys}.

\subsection*{Coarse-grained master equation}
For a given projection $s=\pi(c)$ it is clear that the probabilities of the laboratory system can be obtained by coarse-graining
\begin{equation}
\label{Ps}
P_s(t) \;=\; \sum_{c \in s} P_c(t)\,,
\end{equation}
where the sum is understood to run over all $c$ with $\pi(c)=s$. Similarly, it is clear that the corresponding probability currents sum up as well:
\begin{equation}
\label{currents}
J_{s \to s'}(t) \;=\; \sum_{c \in s} \sum_{c' \in s'} J_{c \to c'}(t)\,.
\end{equation}
Thus, if the total system is assumed to evolve according to the master equation
\begin{equation}
\label{mastereq_tot}
\frac{\d}{\d t} P_c(t)  \;=\; \sum_{c' \in \Omega^{tot} } \Bigl(J_{c' \to c}(t) - J_{c \to c'}(t)\Bigr)\,,
\end{equation}
it is easy to show that these coarse-grained quantities defined in (\ref{Ps})-(\ref{currents}) obey the master equation  (\ref{masterJ}). In other words, a master equation in the total system gives rise to a projected master equation in the embedded laboratory system.

The coarse-grained currents and probabilities allow us to define \textit{effective rates} in the laboratory system:
\begin{equation}
\label{effectiverates}
w_{s\to s'}\left( t \right) \;=\; \frac{J_{s\to s'}\left( t \right)}{P_s\left( t \right)} \,.
\end{equation}
Unlike the rates $w_{c \to c'}$ of the total system, which are time-independent and symmetric, these rates are generally non-symmetric. Moreover, they do not necessarily obey the Kolmogorov criterion, and they may also depend on time~\cite{kolkata,Esposito85}. 

\subsection*{Physical interpretation}

As already mentioned, we assume the total system to be isolated, implying that the transition rates $w_{c \to c'}$ are symmetric. Thus, if the total system is finite and ergodic, it will nevertheless end up in a uniformly distributed state where $J_{c \to c'}=J_{c'\to c}$. This implies that the laboratory system will eventually reach a stationary state obeying detailed balance. 

However, before reaching this stationary state (meaning that the environment is still relaxing towards equilibrium) the laboratory system considered by itself may be found to violate the Kolmogorov criterion, meaning that it is genuinely out of equilbrium. The apparent contradiction that it will end up in a detailed-balanced equilibrium state can be reconciled by observing that the effective transition rates (\ref{effectiverates}) are generally time-dependent: When the total system finally equilibrates the rates change in such a way that detailed balance is restored. 

Thus, if the total system is finite, a possible nonequilibrium dynamics in the laboratory system can only be established for a transient period. However, if the environment is infinite, then it may be possible to keep the laboratory system out of equilibrium for infinite time, allowing the possibility of nonequilibrium steady states.

As we will argue below it is also possible to look at the problem in opposite direction, meaning that for a given nonequilibrium system we can always find a total system with symmetric rates which generates this dynamics. In other words, for a given set of effective rates in the laboratory system we can always 'engineer' an artificial total system which produces them. In the following section we will give an explicit example of such a construction. 

\subsection*{Entropy}
As usual, the entropy of the laboratory and the total system are defined by
\begin{equation}
S^{sys}(t) = - \sum_{s\in\Omega^{sys}} P_s(t) \ln P_s(t),\qquad  
S^{tot}(t) = - \sum_{c\in\Omega^{tot}} P_c(t) \ln P_c(t)\,.
\end{equation}
Using the master equation (\ref{mastereq_tot}) it is straight forward to calculate the entropy production of the laboratory and the total system:
\begin{eqnarray}
\label{ep_schnak}
{\dot S}^{sys}(t)&=&\sum_{s}\sum_{s'} \Bigl( P_{s}(t)w_{s\to s'}(t)-P_{s'}(t)w_{s'\to s}(t) \Bigr)\ln P_{s}(t) \,,\\
\label{ep_total}
{\dot S}^{tot}(t)&=&\sum_{c}\sum_{c'} \Bigl( P_{c}(t)w_{c\to c'}-P_{c'}(t)w_{c'\to c} \Bigr)\ln P_{c}(t) \,.
\end{eqnarray}
The actual entropy production of the total system (\ref{ep_total}) involves the probabilities $ P_c(t) $ and rates $ w_{c\to c'} $ of the total system which are generally not accessible. However, according to Schnakenberg~\cite{Schnakenberg76} we can nevertheless quantify the total entropy production solely on the basis of the probabilities $ P_s(t) $ and the rates $ w_{s\to s'}(t) $ of the laboratory system by means of the formula
\begin{equation}
\label{prod_schnak_curr}
{\dot S}^{tot}_0(t) = \sum_{s,s'}J_{s\to s'}(t)\ln\frac{J_{s\to s'}(t)}{J_{s'\to s}(t)} \,.
\end{equation}
Note that the entropy production according to Schnakenberg (here denoted by the subscript '0') does not necessarily coincide with the real entropy production in~(\ref{ep_total}). This can be seen in the extreme example where the laboratory system consists only of a single configuration. Here ${\dot S}^{tot}_0$ vanishes, whereas ${\dot S}^{tot}$ in Eq. (\ref{ep_total}) may be nonzero. As we will argue below, we expect them to be related by the inequality ${\dot S}^{tot} \geq {\dot S}^{tot}_0$ which reduces to an equality in the limit of instant equilibration in the environment.

\section{Numerical results}\label{sec_num}


\subsection{Simulation of a total system with random properties}

To test the proposed inequality we modeled the total system by an artificial Markov process, where $N_{tot}$ configurations are fully connected by randomly chosen rates. Moreover, we introduced a subsystem by defining an random projection $\pi$ to a smaller set of system $N_{sys}$ configurations. Solving the master equation of the system numerically we compute the corresponding entropies and study the entropy production.

The numerical analysis was implemented as follows. A vector of size $ N_{tot} $ was set up containing the probabilities $ P_c(t) $ of the total system.
This array was then filled with random numbers and normalized, defining the initial probability distribution of the total system.
Then every entry was randomly assigned to a configuration $ s $ of the laboratory system.
By summing over all probabilities $ P_c(t) $ mapped to the same $s$ the probability $ P_s(t) =\sum_{c\in s} P_c(t)$ can be determined.
Finally the setup was completed by generating random symmetric rates $ w_{c\to c'}=w_{c'\to c} $ between all array entries.

\begin{figure}
\centering\includegraphics[width=150mm]{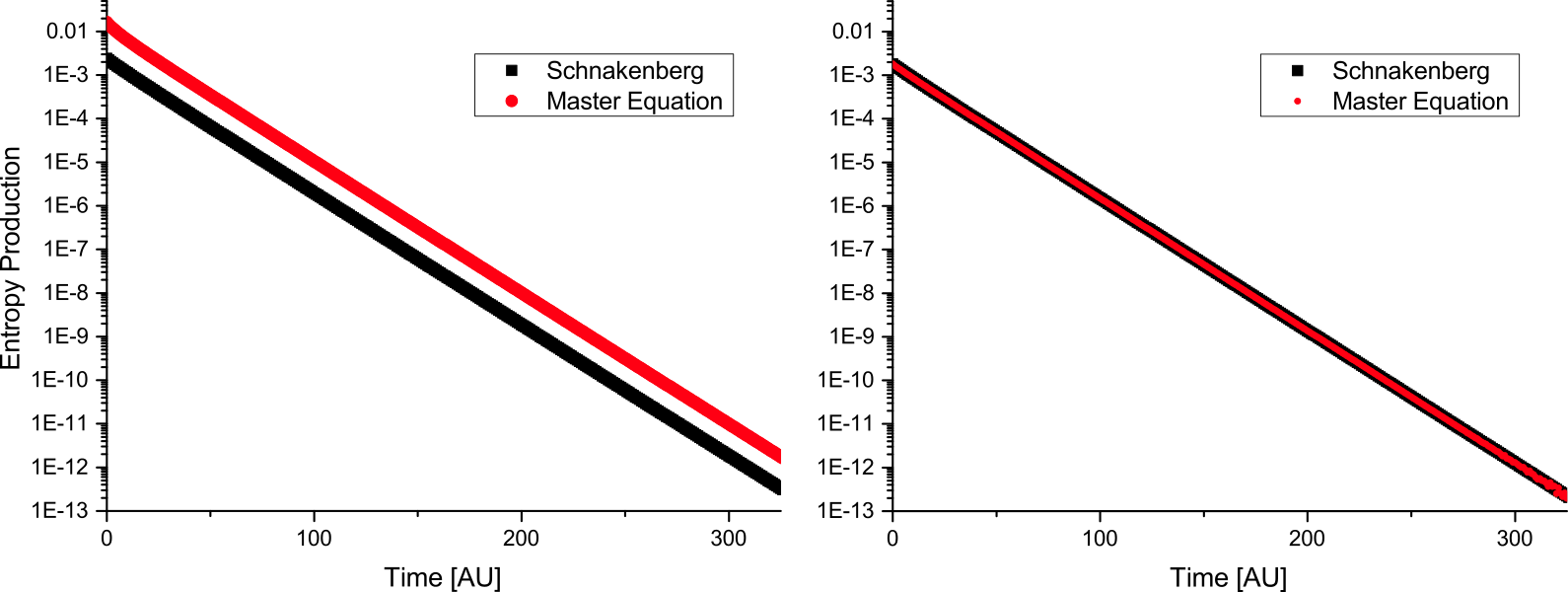}
\caption{Left: Temporal evolution of the actual entropy production (red) and the prediction by Schnakenberg (black) in a random Markov process with $N_{tot}=1000$ configurations mapped to $N_{sys}=200$ system configurations. Right: If one adds an artificial instant equilibration of the environment after each update, the two curves coincide.}
\label{figschnakenberg}
\end{figure}

Solving the master equation with a vectorized Runge-Kutta algorithm  we computed the Shannon entropy of the total system before and after each time step. The change of this entropy is considered as the actual total entropy produced in the time interval. As the Schnakenberg formula generates a value for the time derivative of the entropy, the corresponding entropy production was calculated via the trapezoidal rule.

As can be seen in Fig.~\ref{figschnakenberg}, the produced entropy for this random system is obviously at all times larger than the Schnakenberg entropy production. In the completely random initial state there is a difference which decreases as the system relaxes towards equilibrium. We repeated this calculation under various conditions, obtaining qualitatively similar results. This suggests that the inequality
\begin{equation}
\label{inequality}
{\dot S}^{tot}(t) \geq {\dot S}^{tot}_0(t)
\end{equation}
holds in any system, i.e., the Schnakenberg formula provides a lower bound of the actual entropy production.

In a second numerical study we equilibrated all subsectors of the total system instantly after each update, meaning that the $P_c(t)$ for all $c$ belonging to the same system configuration $s$ were forced to coincide. As shown in the right panel of Fig.~\ref{figschnakenberg} one obtains a perfect coincidence. This suggests that the inequality becomes sharp in the limit of instant equilibration.
We will come back to this point in Section \ref{sec_analytical}.

\subsection{Construction of the environment of a clock model}

\begin{figure}
\centering\includegraphics[width=35mm]{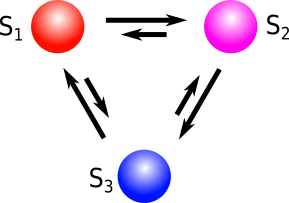}
\caption{Cyclic clock model with three configurations and asymmetric rates. The process jumps preferably in clockwise direction, leading to a non-vanishing probability current even in the stationary state.}
\label{figclock}
\end{figure}

In order to demonstrate that for a given nonequilibrium system we can 'engineer' an appropriate embedding into a larger environment, we consider a class of cyclic models, where the configurations are connected with asymmetric rates. The simplest case is a three-state clock model, as shown in Fig.~\ref{figclock}.

\begin{figure}
	\centering\includegraphics[width=70mm]{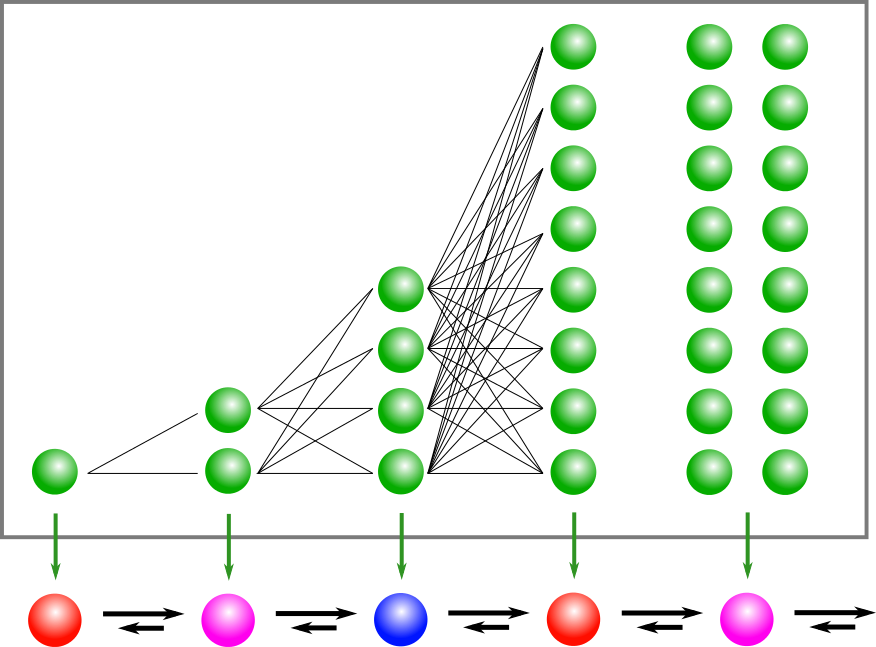}
	\caption{Interpretation of the cyclic process as a biased random walk on a linear chain. The upper box shows the corresponding configurations of the total system (see text).}
	\label{figclock_total}
\end{figure}

Since the clock model does not obey detailed balance, it is obviously out of equilibrium. In the following we show that this dynamics can be generated by embedding it into a larger equilibrium system, i.e., we create a suitable model of the environment in such a way that the total system is isolated, having only symmetric rates. To this end we first unravel the cycle into a linear chain of repeating configurations~\cite{kolkata}, as sketched in Fig.~\ref{figclock_total}. As can be seen, each configuration on the chain corresponds to a number of configurations $c$ in the total system, which means there are new three levels to view this model, namely,
\begin{itemize}
 \item the level of the clock model  having only three configurations, 
 \item the corresponding linear chain of repeating configurations (unraveled states), and 
 \item the total system which allows for a large number of configurations for each unraveled state.
\end{itemize}
On the linear chain the model performs a random walk that is biased to the right. In order to generate this dynamics, only neighboring sets of configurations in the environment are connected, as sketched in the Fig.~\ref{figclock_total} (the configurations within each column may be connected as well, which is not shown). Note that the configurations of the total system are all connected by \textit{symmetric} rates. Thus, in order to create a bias on the linear chain, the number of configurations in the total system has to increase as we go to the right. In this way an entropic force will be introduced, dragging the system to the right.

Enumerating the states on the linear chain by an index $i$, the ratio of the rates
\begin{equation}
r=\frac{w_\rightarrow}{w_\leftarrow}
\end{equation}
determines how quickly the number $n_i$ of configurations in each columns grows. For example, for $r=2$ this number doubles from column to column, increasing exponentially as $n_i=2^{i-1}$. Moreover, the symmetric rates in the total system have to be chosen in such a way that the probability current along the chain remains constant. This also means that on the chain the symmetric rates of the total system have to be proportional to $1/n_i$. 

If the linear chain was infinite with an ever-increasing $n_i$, then the projected clock model would indeed rotate forever. However, on a finite chain the system will eventually reach the right edge. When the edge is reached the effective rates of the laboratory system will change, leaving the non-equilibrium steady state and establishing detailed balance.

\subsection{Solving the master equation of a clock model numerically}

In order to avoid the exponentially increasing number of configurations in the total system in a numerical analysis, we use the symmetry property that all configurations in a given column are equally probable. Therefore, it suffices to keep track of only one of them. In the end this probability has to be reweighted by $n$.

We solved this problem numerically on a chain with $N=500$ columns, starting with an initial probability distribution focused in the center. The numerical solution is then used to determine the actual entropy production and to compare it with the prediction by Schnakenbergs formula calculated on the level of the clock model.

\begin{figure}
	\centering\includegraphics[width=80mm]{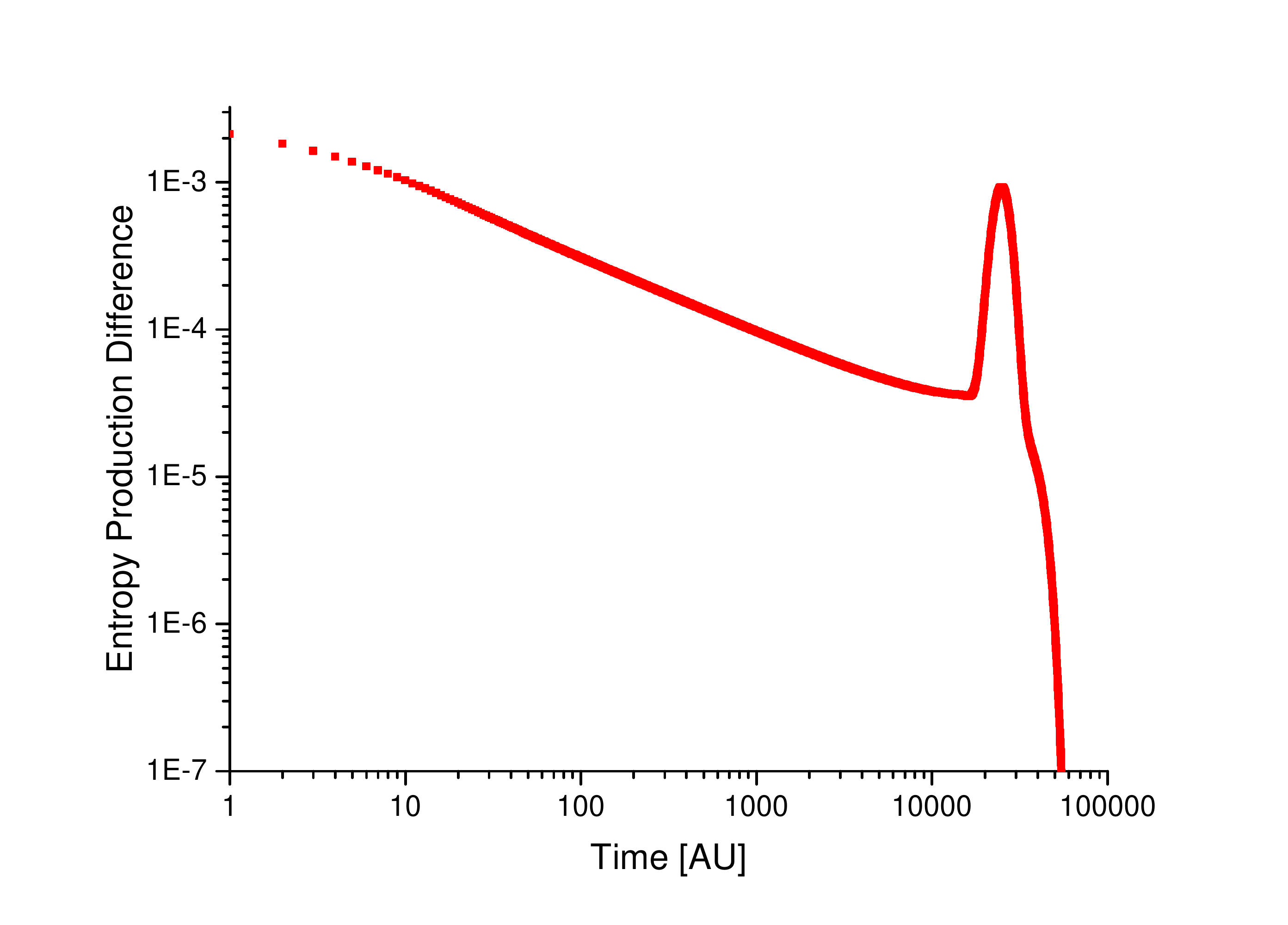}
	\caption{Temporal evolution of the difference between actual entropy production and the prediction by Schnakenberg in cyclic model with $3$ configurations and the rates $w_\rightarrow=2$ and $w_\leftarrow=1$, modeled as a chain with length $N=500$.}
	\label{figclockdiff}
\end{figure}

The difference between the two quantities is plotted in Fig.~\ref{figclockdiff}. As can be seen, the difference first decreases like $\sqrt{t}$, which is expected for a diffusion-like process. The diffusion-like behavior can be explained by recalling the three different levels of description:
\begin{itemize}
\item The Schnakenberg formula refers to the three-state configuration space of the laboratory system. After a short initial transient the configuration entropy becomes constant $(H_{sys} = \ln 3)$ and the entropy production in the environment saturates at a constant value $\dot S=\frac 12 \ln 2$.  
\item Contrarily, the actual entropy production refers to the level of the total system. It consists of two contributions. On the one hand the biased motion to the right contributes with a constant entropy production. On the other hand, the probability distribution broadens, giving an additional contribution to the entropy production. This contribution is expected to scale as $1/\sqrt{t}$. This explains the straight slope in the double-logarithmic plot in Fig.~\ref{figclockdiff}.
\end{itemize}
Another notable feature is the pronounced peak in the figure which can be identified with the moment when the right edge of the linear chain is reached. Here the effective rates in the laboratory system start to change, leading to a sudden increase of the difference. As the system is further relaxing towards detailed balance both entropy productions are decreasing again which accounts for the final rapid decay after the peak. Throughout the whole time evolution the Schnakenberg entropy production is smaller than the actual one, confirming that it provides a lower bound.

\section{Analytical results\label{sec_analytical}}

In order to support that the Schnakenberg entropy production is in fact a lower bound, we proceed in two steps. First we prove that in the case of instant equilibration of the environment the Schnakenberg entropy production and the actual entropy production coincide. Then we prove that the Schnakenberg entropy production provides a lower bound to the actual entropy production. In the following sections we omit the time-dependencies of the variables introduced in Sect. \ref{sec_analytical} for ease of reading.

\subsection*{Special case of instant equilibration in the environment}

Since the total system is isolated, the rates $w_{c \to c'}$ are symmetric. This allows the actual total entropy production to be rewritten as
\begin{eqnarray}
\label{prod_tot_rewr}
    {\dot S}^{tot} & = & \sum_{c\in \Omega^{tot}}\sum_{c'\in \Omega^{tot}} \left( P_{c}w_{c\to c'}-P_{c'}w_{c'\to c} \right)\ln P_{c} \nonumber \\
            & = & \sum_{c\in \Omega^{tot}}\sum_{c'\in \Omega^{tot}} P_{c}w_{c\to c'} \ln P_{c} - \sum_{c\in \Omega^{tot}}\sum_{c'\in \Omega^{tot}}P_{c'}w_{c'\to c} \ln P_{c} \nonumber \\
            & = & \sum_{c\in \Omega^{tot}}\sum_{c'\in \Omega^{tot}} P_{c}w_{c\to c'} \ln P_{c} - \sum_{c'\in \Omega^{tot}}\sum_{c\in \Omega^{tot}}P_{c}w_{c\to c'} \ln P_{c'} \nonumber \\
            & = & \sum_{c\in \Omega^{tot}}\sum_{c'\in \Omega^{tot}} P_{c}w_{c\to c'} \ln \frac{P_{c}}{P_{c'}}\,.  
\end{eqnarray}
Using the map to the laboratory system the sums can be reorganized by
\begin{eqnarray}
\label{sum}
    {\dot S}^{tot} & = & \sum_{s\in \Omega^{sys}}\sum_{s'\in \Omega^{sys}}\sum_{c\in s}\sum_{c'\in s'} P_{c}w_{c\to c'} \ln \frac{P_{c}}{P_{c'}}.
\end{eqnarray}
Now let us assume that the environment equilibrates instantly, meaning that all probabilities $P_c$ belonging to the same sector $s$ coincide, i.e.
\begin{equation}
 P_{c}=p_{s}  \qquad \forall c \in s\,,
\end{equation}
where $p_s$ should not be confused with $P_{s}=\sum_{c\in s}P_{c}$. Under this assumption Eq.~(\ref{sum}) reduces to
\begin{eqnarray}
    {\dot S}^{tot} & = & \sum_{\stackrel{s,s'\in \Omega^{sys}}{  s\neq s'}}
    p_s \ln \frac{p_{s}}{p_{s'}}
    \sum_{c\in s}\sum_{c'\in s'} w_{c\to c'} ,
    \label{Shannon_simp_eq}
\end{eqnarray}
where the first sum runs only over different sectors (columns) because of the logarithm.

Now we show that the Schnakenberg entropy production (\ref{ep_schnak}) leads to the same expression. Using Eq.~(\ref{currents}) we get
\begin{eqnarray}
    {\dot S}^{tot}_0 & = & \sum_{\stackrel{s,s'\in \Omega^{sys}}{  s\neq s'}}
    J_{s\to s'}\ln\frac{J_{s\to s'}}{J_{s'\to s}} \nonumber \\
              & = & \sum_{\stackrel{s,s'\in \Omega^{sys}}{  s\neq s'}} \left( \sum_{c\in s}\sum_{c'\in s'} P_{c}w_{c\to c'} \right) \ln \frac{\sum_{c\in s}\sum_{c'\in s'} P_{c} w_{c\to c'}}{\sum_{c\in s}\sum_{c'\in s'} P_{c'} w_{c'\to c}} \,. 
\end{eqnarray}
By using the assumption of instant equilibration of the environment and the symmetry of the rates in the total system we can further simplify this formula:
\begin{eqnarray}
    {\dot S}^{tot}_0 & = & \sum_{\stackrel{s,s'\in \Omega^{sys}}{  s\neq s'}}\left( \sum_{c\in s}\sum_{c'\in s'} p_{s} w_{c\to c'} \right) \ln \frac{ p_{s} \sum_{c\in s}\sum_{c'\in s'} w_{c\to c'}}{ p_{s'} \sum_{c\in s}\sum_{c'\in s'} w_{c'\to c}} \nonumber \\
              & = &\sum_{\stackrel{s,s'\in \Omega^{sys}}{  s\neq s'}}
              p_{s}  \ln \frac{ p_{s}}{p_{s'}}
              \sum_{c\in s}\sum_{c'\in s'} w_{c\to c'} \label{Schnak_simp_eq}
\end{eqnarray}
If we now compare (\ref{Schnak_simp_eq}) to (\ref{Shannon_simp_eq}), we see, that for instant equilibration ($ P_{c}=P_{c'}=p_s $ for $ c,c'\in s $) and symmetric rates in the total system the actual entropy production calculated using the Schnakenberg formula is equal to the entropy production calculated via Shannon formula and the master equation.

\subsection*{Schnakenberg formula as the minimal entropy production}

Now we prove that the Schnakenberg entropy production provides a lower bound to the actual entropy production. To this end we first define the function
\begin{equation}
\label{hxy}
h(x,y)=(x-y)\ln \frac{x}{y},
\end{equation}
with $x,y\in \mathbb{R}^+$. For $x\neq y$ this function is positive and zero otherwise. The Hessian matrix of $h$ 
\begin{eqnarray}
\mathbf{H} & = &
\left(\begin{array}{cc}
\frac{\partial^2 h(x,y)}{\partial x^2} & \frac{\partial^2 h(x,y)}{\partial x\partial y} \\
\frac{\partial^2 h(x,y) }{\partial y\partial x} & \frac{\partial^2 h(x,y) }{\partial y^2}
\end{array}\right) \;=\;
\left(\begin{array}{cc}
\frac{x+y}{x^2} & -\frac{x+y}{xy} \\
-\frac{x+y}{xy} & \frac{x+y}{y^2}
\end{array}\right)
\end{eqnarray}
is positive semidefinite. This implies that $h$ is convex, which in turn allows us to use Jensen's inequality:
\begin{equation}
\label{Jensen}
\sum_{i=1}^{n}a_ih(x_i,y_i)\geq h\left(\sum_{i=1}^{n}a_ix_i,\sum_{i=1}^{n}a_iy_i\right),\qquad a_i\geq 0,\qquad \sum_{i=1}^{n}a_i.
\end{equation}
Furthermore one can easily see that $h$ is homogenous, i.e.,
\begin{equation}
\label{extract}
h\left(\frac{x}{\lambda},\frac{y}{\lambda}\right)=\frac{h(x,y)}{\lambda}\,.
\end{equation}
To describe the entropy production with Eq. (\ref{hxy}) we need to rewrite Eq. (\ref{prod_tot_rewr}) again, using the symmetry of the rates in the total system and the definition of the probability currents:
\begin{eqnarray}
    {\dot S}^{tot} & = & \sum_{c\in \Omega^{tot}}\sum_{c'\in \Omega^{tot}} P_{c}w_{c\to c'} \ln \frac{P_{c}}{P_{c'}} \nonumber \\
                   & = & \sum_{c\in \Omega^{tot}}\sum_{c'\in \Omega^{tot}} P_{c}w_{c\to c'} \ln \frac{P_{c}w_{c\to c'}}{P_{c'}w_{c'\to c}} \nonumber \\
                   & = & \sum_{c\in \Omega^{tot}}\sum_{c'\in \Omega^{tot}} J_{c\to c'} \ln \frac{J_{c\to c'}}{J_{c'\to c}}
\end{eqnarray}
This can be further rewritten by switching indices providing us with the desired form:
\begin{eqnarray}
\label{tot_h}
    {\dot S}^{tot} & = & \sum_{c\in \Omega^{tot}}\sum_{c'\in \Omega^{tot}} \frac{1}{2} (J_{c\to c'}+J_{c\to c'}) \ln \frac{J_{c\to c'}}{J_{c'\to c}} \nonumber \\
                   & = & \frac{1}{2} \left(
                   \sum_{c\in \Omega^{tot}}\sum_{c'\in \Omega^{tot}} J_{c\to c'} \ln \frac{J_{c\to c'}}{J_{c'\to c}} + 
                   \sum_{c\in \Omega^{tot}}\sum_{c'\in \Omega^{tot}} J_{c\to c'} \ln \frac{J_{c\to c'}}{J_{c'\to c}}\right) \nonumber \\
                   & = & \frac{1}{2} \left(
                   \sum_{c\in \Omega^{tot}}\sum_{c'\in \Omega^{tot}} J_{c\to c'} \ln \frac{J_{c\to c'}}{J_{c'\to c}} + 
                   \sum_{c'\in \Omega^{tot}}\sum_{c\in \Omega^{tot}} J_{c'\to c} \ln \frac{J_{c'\to c}}{J_{c\to c'}}\right) \nonumber \\
                   & = & \frac{1}{2} \left(
                   \sum_{c\in \Omega^{tot}}\sum_{c'\in \Omega^{tot}} J_{c\to c'} \ln \frac{J_{c\to c'}}{J_{c'\to c}} - 
                   \sum_{c'\in \Omega^{tot}}\sum_{c\in \Omega^{tot}} J_{c'\to c} \ln \frac{J_{c\to c'}}{J_{c'\to c}}\right) \nonumber \\
                   & = & \frac{1}{2} \sum_{c\in \Omega^{tot}}\sum_{c'\in \Omega^{tot}} (J_{c\to c'}-J_{c'\to c}) \ln \frac{J_{c\to c'}}{J_{c'\to c}} \nonumber \\
                   & = & \frac{1}{2} \sum_{c\in \Omega^{tot}}\sum_{c'\in \Omega^{tot}} h(J_{c\to c'},J_{c'\to c})
\end{eqnarray}
Rewriting Eq.~(\ref{prod_schnak_curr}) provides us with a similar result for the Schnakenberg entropy production:
\begin{eqnarray}
\label{schnak_h}
{\dot S}^{tot}_0 & = & \sum_{s\in \Omega^{sys}}\sum_{s'\in \Omega^{sys}}J_{s\to s'}\ln\frac{J_{s\to s'}}{J_{s'\to s}} \nonumber \\
                 & = & \sum_{s\in \Omega^{sys}}\sum_{s'\in \Omega^{sys}}\frac{1}{2}(J_{s\to s'}+J_{s\to s'})\ln\frac{J_{s\to s'}}{J_{s'\to s}} \nonumber \\
                 & = & \frac{1}{2}\left(
                 \sum_{s\in \Omega^{sys}}\sum_{s'\in \Omega^{sys}}J_{s\to s'}\ln\frac{J_{s\to s'}}{J_{s'\to s}} + 
                 \sum_{s\in \Omega^{sys}}\sum_{s'\in \Omega^{sys}}J_{s\to s'}\ln\frac{J_{s\to s'}}{J_{s'\to s}}\right) \nonumber \\
                 & = & \frac{1}{2}\left(
                 \sum_{s\in \Omega^{sys}}\sum_{s'\in \Omega^{sys}}J_{s\to s'}\ln\frac{J_{s\to s'}}{J_{s'\to s}} + 
                 \sum_{s'\in \Omega^{sys}}\sum_{s\in \Omega^{sys}}J_{s'\to s}\ln\frac{J_{s'\to s}}{J_{s\to s'}}\right) \nonumber \\
                 & = & \frac{1}{2}\left(
                 \sum_{s\in \Omega^{sys}}\sum_{s'\in \Omega^{sys}}J_{s\to s'}\ln\frac{J_{s\to s'}}{J_{s'\to s}} - 
                 \sum_{s'\in \Omega^{sys}}\sum_{s\in \Omega^{sys}}J_{s'\to s}\ln\frac{J_{s\to s'}}{J_{s'\to s}}\right) \nonumber \\
                 & = & \frac{1}{2}\sum_{s\in \Omega^{sys}}\sum_{s'\in \Omega^{sys}}(J_{s\to s'}-J_{s'\to s})\ln\frac{J_{s\to s'}}{J_{s'\to s}} \nonumber \\
                 & = & \frac{1}{2}\sum_{s\in \Omega^{sys}}\sum_{s'\in \Omega^{sys}}h(J_{s\to s'},J_{s'\to s})
\end{eqnarray}
We now assume that the set $\pi^{-1}(s)=\{c:\pi (c)=s\}$ is finite for all $s$ and let $n_s$ be the number of total system configurations $c$ of which each laboratory configuration $s$ consists. Using Eq.~(\ref{Jensen}) with $n=n_s\cdot n_{s'}$, $a_i=\frac{1}{n}$ and $x_i=J_{c\to c'}$, $y_i=J_{c'\to c}$ and summing over $c\in s$ and $c'\in s'$ we get the following inequality:
\begin{equation}
\frac{1}{n}\sum_{c\in s}^{n_s}\sum_{c'\in s'}^{n_{s'}}h(J_{c\to c'},J_{c'\to c})
\geq h\left(\frac{1}{n}\sum_{c\in s}^{n_s}\sum_{c'\in s'}^{n_{s'}}J_{c\to c'},\frac{1}{n}\sum_{c\in s}^{n_s}\sum_{c'\in s'}^{n_{s'}}J_{c'\to c}\right).
\end{equation}
Taking $\frac{1}{n}$ out from the right hand side using Eq.~(\ref{extract}) and further simplifying by using Eq. (\ref{currents}) we obtain the following inequality:
\begin{eqnarray}
\label{ineq_2}
\frac{1}{n}\sum_{c\in s}^{n_s}\sum_{c'\in s'}^{n_{s'}}h(J_{c\to c'},J_{c'\to c})
& \geq & \frac{1}{n}h\left(\sum_{c\in s}^{n_s}\sum_{c'\in s'}^{n_{s'}}J_{c\to c'},\sum_{c\in s}^{n_s}\sum_{c'\in s'}^{n_{s'}}J_{c'\to c}\right) \nonumber \\
\phantom{\frac{1}{n}}\sum_{c\in s}^{n_s}\sum_{c'\in s'}^{n_{s'}}h(J_{c\to c'},J_{c'\to c})
& \geq & h\left(\sum_{c\in s}^{n_s}\sum_{c'\in s'}^{n_{s'}}J_{c\to c'},\sum_{c\in s}^{n_s}\sum_{c'\in s'}^{n_{s'}}J_{c'\to c}\right) \nonumber \\
& = & h(J_{s\to s'},J_{s'\to s}).
\end{eqnarray}
We can now apply the sum over all $s,s'\in \Omega^{sys}$ and the factor $\frac{1}{2}$ to both sides of Eq. (\ref{ineq_2}) and compare it to Eq. (\ref{tot_h}) and Eq. (\ref{schnak_h}), which will result in the desired inequality:
\begin{eqnarray}
\frac{1}{2}\sum_{s\in \Omega^{sys}}\sum_{s'\in \Omega^{sys}}\sum_{c\in s}^{n_s}\sum_{c'\in s'}^{n_{s'}}h(J_{c\to c'},J_{c'\to c})
& \geq & \frac{1}{2}\sum_{s\in \Omega^{sys}}\sum_{s'\in \Omega^{sys}}h(J_{s\to s'},J_{s'\to s}) \nonumber \\
{\dot S}^{tot} & \geq & {\dot S}^{tot}_0 
\end{eqnarray}
This completes the proof that the Schnakenberg entropy production provides a lower to the actual entropy production.

\section{Conclusions and outlook}

In this work we have tried to explain entropy production of nonequilibrium systems in the environment by embedding the laboratory system into a larger isolated total system without introducing the notions of energy conservation, work, heat and temperature. We have shown that the commonly accepted formula for this purpose found by Schnakenberg is only exact in the case of instant equilibration of the environment. This is not surprising since Schnakenberg himself derived this formula by using methods of equilibrium thermodynamics, implicitly assuming that the environment is arbitrarily close to equilibrium. 

The conjecture that the Schnakenberg formula provides a lower bound of the actual entropy production of the total system was proven in general and confirmed by a numerical investigation of a random Markov process. Furthermore we have given an analytical argument showing that the actual entropy production becomes minimal, and thereby equal to the one given by the Schnakenberg formula, in the case of infinitely fast equilibration among the corresponding configurations in the environment. We have tested these findings numerically by creating an artificial environment for a simple three-state nonequilibrium clock model.\\

\noindent
\textbf{Acknowledgments}\\
HH would like to thank the Galileo Galilei Institute for Theoretical Physics for the hospitality and the INFN for partial support during the completion of this work.\\

\end{document}